\documentclass{elsarticle}

\usepackage{graphicx}
\usepackage{graphicx}
\usepackage{caption}
\usepackage{epstopdf}
\usepackage{epsfig}
\usepackage[round]{natbib}

 \usepackage{mathptmx}       
\usepackage{amsmath,amssymb,eucal,upref,bm}
\usepackage{graphicx,amsmath,amssymb,fixmath}

\newdefinition{rmk}{Remark}
\newproof{pf}{Proof}
\newproof{pot}{Proof of Theorem \ref{thm2}}

\begin{document}

\title{Large ecosystems in transition: \\ bifurcations and mass extinction}

\author[rvt]{Ivan Sudakov\corref{cor1}}
\ead{isudakov1@udayton.edu}
\author[focal,els]{Sergey A. Vakulenko} 
\author[els]{Dubrava Kirievskaya} 
\author[rvt4]{Kenneth M. Golden}

\cortext[cor1]{Corresponding author}
\address[rvt]{University of Dayton, Department of Physics, 300 College Park, SC 111, Dayton, OH 45469-2314 USA}
\address[focal]{Institute of Problems in Mechanical Engineering, Russian Academy of Sciences, Bolshoy pr., 61, V.O., St.\,Petersburg 199178, Russia}
\address[els]{University ITMO, Kronverkskiy pr., 49, St.\,Petersburg 197101, Russia}
\address[rvt4]{University of Utah, Department of Mathematics, 155 S 1400 E RM 233,
Salt Lake City, UT 84112-0090, USA}

\begin{abstract}

We propose a model of multispecies populations surviving on  
distributed resources. System dynamics 
are investigated under changes in
abiotic factors such as the climate, 
as parameterized through environmental temperature. 
In particular, we 
introduce a feedback between species abundances and 
resources via abiotic factors. This model is apparently 
the first of its kind to include a feedback mechanism 
coupling climate and population dynamics.
Moreover, we take into 
account self-limitation effects. The model 
explains the coexistence of many species, yet also displays the 
possibility of catastrophic bifurcations, where all 
species become extinct under the influence of abiotic factors. 
We show that as these factors change there are different 
regimes of ecosystem behavior, including a possibly
chaotic regime when abiotic influences are sufficiently strong.

\end{abstract}


\maketitle
%
\section{Introduction}

Models of ecosystems form an important class of dynamical systems generating complex dynamics, bifurcations and strange  attractors \citep{Ula79}. 
However, modeling these large systems is made difficult by rapid, large scale biological evolution and gaps in observations to use for comparison. 
Also, there is uncertainty in how to set up reliable 
experiments on such ecosystems. 

Recent observations have shown that climate change may be a leading factor 
influencing ecosystem behavior \citep{Wal2010}. Large multispecies marine ecosystems are sensitive indicators of climate change \citep{Greb11, Ked}. 
As a key part of the global ecosystem, they influence climate 
feedback processes and possible tipping points \citep{Sel2015}. 
A well studied example 
is the ocean ecosystem, where 
phytoplankton are the main resource
for many species.    
Phytoplankton populations play an important role in the dynamics 
of the climate system through the 
oceanic carbon cycle -- by 
removing about half of all carbon dioxide from the atmosphere 
during photosynthesis \citep{FiBeRaFa98}. Previous 
studies \citep{ArBre04, TraShJenCu07} have shown that 
phytoplankton communities respond to climate warming through
changes in diversity and productivity. However, it was 
recently determined \citep{Tos13} that changing the 
climate temperature directly impacts the 
chemical cycles in 
plankton, affecting the system as much as 
nutrients and light.

We consider here a model of a large ecosystem where many species share few resources. 
It extends the model of phytoplankton species competition in \citep{HuWe99}, by
taking into account that the resources depend
on environmental factors,
in particular, climate, as well as self-limitation
and competition effects.  
Our aim is to explore the connections among 
complexity of the temporal behavior, 
biodiversity, and the structure of 
the climate--ecosystem interaction.

Note that competition may occur as a result of  
the following mechanism \citep{Roy}.
There are a number of species of phytoplankton which
have the ability to produce some toxic or inhibitory compounds.
These toxic materials compensate for the
competitive disadvantages among phytoplankton species which 
leads to self-limitation effects. Moreover, resource levels  
may depend on the environment via temperature or greenhouse gas concentration. 

Many mathematical models \citep{Hof88, Ta96, Zeem} show
that only a single species can survive in an ecosystem for certain
fixed parameters. Biologically, 
this is the competitive exclusion principle. 
In the framework of the phytoplankton 
model, it is known as the so-called {\it plankton paradox} 
studied in many interesting 
works \citep{Hu61,Til77,HuWe99,Hu2005}.
In particular, it is sometimes observed in nature that 
numerous species can coexist while
depending on the same resource, even though
competition tends to exclude species.
In fact, in contrast to the exclusion principle,
we observe here the coexistence of many plankton species 
sharing the same niche and resources.
Numerical simulations \citep{Hu61,Til77,HuWe99} 
have shown that in such systems 
chaos and unpredictable behavior occur.  
In \citep{Hsu,Smith81} it was shown that temporal variability 
of the nutrient supply can lead to coexistence of species.



The environment may alter the distribution and 
abundance of the species in a population. Such 
effects have been studied in terms of internal 
processes within the population, like competition 
for resources and
conditions for chemical reactions. 
However, current models have not been linked 
to feedback with the environment. 
Feedback between a population and the
environment can occur as a result of changes 
in abiotic factors such as temperature, nutrient 
concentrations, and light intensity. 

The main results of this paper show that the 
population dynamics depends sharply on 
feedback with the environment. 
For simplicity, hereinafter we refer to this  
as \textit{climate -- ecosystem feedback}. 
If the abiotic factor is temperature $T$, 
for example, then it is natural 
to talk about the feedback between an ecosystem 
and the climate system, which can be parameterized as a function 
of a rate of change of the resource supply with 
respect to temperature. If that feedback is 
negative -- where species abundance decreases resources -- 
then an ecosystem can support a number of species and 
the dynamics is relatively simple (non-chaotic and non-periodic). 
If the feedback is positive -- where species abundance 
increases resources -- then for a sufficiently 
large feedback level there are possible mass extinctions  
which occur suddenly, and moreover, there are possible
chaotic or periodic dynamics.

The paper is organized as follows. In the next section we formulate 
the standard model of species coexistence and the extended model, 
which takes into account climatic factors.
Further, in section \ref{Genprop} we prove a general assertion 
on the existence of an attractor for this model. 
In section \ref{Asym} it is shown that for 
large turnover rates $D$ the  
system admits an asymptotic solution and, 
under additional assumptions, 
can be reduced to the Lotka-Volterra model \citep{KoVa, Koz}. 
This model is well studied \citep{Hof88, Ta96, Zeem, Zeem2} 
and known results allow us to describe the 
influence of climate and climate warming
in large ecosystems (see section \ref{Qual}).  
In section \ref{eq}, for the case of a single resource, we  
show that the global attractor consists of equilibria and derive 
an equation for the species abundances. This investigation is 
aimed at describing the influence of climate on biodiversity.

\section{Models of large ecosystems } \label{Model}

\subsection{Standard model}

Consider the following  model of an 
ecosystem with $N$ species, which extends the model of 
resource competition in \citep{HuWe99}:
\begin{equation}
     \frac{dx_i}{dt}=x_i (- r_i  + \phi_i(v) - 
     \sum_{j=1}^N \gamma_{ij} \; x_j),
     \;\;\;\;\;\; 1 \leq i \leq N,
    \label{HX11}
     \end{equation}
\begin{equation}
     \frac{dv}{dt}=D(S-v)   -  \sum_{j=1}^N c_j \; x_j \; \phi_j(v),
    \label{HV11}
     \end{equation}
where
\begin{equation}
      \phi_j(v)= \frac{a_j v}{K_j +v}\;, \quad  a_j , \ K_j >0, 
\label{MM5}
     \end{equation}
is the specific growth rate of species $j$ as a
function of the availability $v$ of the resource 
(also known as Michaelis-Menten's function),  
$x_i$ are species abundances,
$r_i$ are the species moralities, 
$D$ is the resource turnover rate, 
$S$ is the supply concentration of the resource, 
and $DS$ can be interpreted as the supply rate.
The dynamics of the species depend on the 
availability of the resource, which in turn
depends on the rate of resource 
supply and the amount of resource used by the species.

The coefficient $c_j$ is
the content of the resource 
in the $i$-th species. The constants  $c_j$
define how different species share resources. 
Note that if all  $c_j=0$ then
the equation for $v$ becomes trivial and 
$v(t) \to S$ for large times $t$, i.e., the resource
equals the resource supply. We consider this system in the non-negative cone: $x \in {\bf R}_{+}^N$, 
$v >0$, where ${\bf R}^N_{+} =\{x:  x_j \ge 0, \ \forall j\}$. 
The coefficients $a_i$ are 
specific growth rates and the $K_i$ are self-saturation constants.    

We assume that the $\gamma_{ii} >0$.  The terms $\gamma_{ii}x_i$ define
self-regulation of species populations that
restricts their abundances.  In the case $\gamma_{ij} >0$  
 with $i \ne j$ these terms describe competition between species. 
These effects can appear as a result of  an ability to produce some toxic or inhibitory compounds \citep{Roy}.   However, we admit the possibility of mutualistic interactions, in which case $\gamma_{ij} <0$.
Assumptions on $\gamma_{ij}$ are formulated below, 
at the beginning of Section 3.

 
For the case of $M$ resources, we have the more complicated equations
\begin{equation}
     \frac{dx_i}{dt}=x_i (- r_i  + \phi_i(v) -  
     \sum_{k=1}^N \gamma_{ik} \; x_k), \;\;\;\;\;\; 1 \leq i \leq N,
    \label{HX2}
     \end{equation}
\begin{equation}
     \frac{dv_j}{dt}=D_j(S_j -v_j)   -  
     \sum_{k=1}^N c_{jk} \; x_k  \; \phi_k(v), \;\;\;\;\;\; 1 \leq j \leq M,
    \label{HV2}
     \end{equation}
where $v=(v_1, v_2, ..., v_M)$,   and the $\phi_j(v)$ are smooth functions.  
We  consider 
 general $\phi_j$  satisfying  the conditions
\begin{equation}
      \phi_j(v) \in C^1, \quad    0 \le \phi_j(v) \le  C_+,  
\label{MM2a}
     \end{equation}
where $C_{+} >0$ is a positive  constant, and
\begin{equation}
      \phi_k(v) =0,   \quad  \forall k, \quad v \in \partial {\bf R}^M_{+},     
\label{MM2b}
     \end{equation}
where $\partial {\bf R}^M_{+} $ denotes the boundary of the  cone $ {\bf R}_{+}^M=\{v:  v_j \ge 0, \ \forall j \}$.
Condition (\ref{MM2a}), in particular, means that $C_{+}$ forms a
uniform upper bound for the $\phi_j(v)$. 
We  assume that
$
 c_{jk} >0.
$
This model is widely used for primary producers 
like phytoplankton, and can also 
be applied to describe competition for terrestrial plants 
\citep{Til77}.

When $\gamma_{ij}=0$ for all $i , j$ this system is 
equivalent to those in works 
where the plankton paradox is studied \citep{HuWe99}.  
The choice $\gamma_{ii} =\gamma_i > 0$ and 
$\gamma_{ij}=0$ for $i \ne j$
allows us to take into account 
self-limitation effects, which is important in 
these systems, as shown by \citet{Roy}.

Below we use the notation $f_{+}=\max\{f, 0\}$. 
We define the scalar product in ${\bf R}^N$ together
with the corresponding norm by
\begin{equation} \label{notat}
\langle f,  g  \rangle_C = \sum_{j=1}^N C_j f_j g_j, \quad  ||f||^2_C=\langle f,f \rangle_C.   
\end{equation}
 This scalar product is defined for $N$-component vectors and depends 
on non-negative coefficients $C_j >0$, $j=1,..., N$.

\subsection{Extended standard model with climate influence}

We extend the system (\ref{HX2}) and (\ref{HV2}) 
to describe potential effects connected with an influence 
of the climate. In fact, temperature has a 
significant effect on the maximum
growth rate of phytoplankton \citep{Richardson}, and can be
considered as a crucial factor in population dynamics.

For one and two species ($N=1,2$),
a model of climate influence was proposed 
in \citep{Petr}. We consider the case of 
arbitrary $N$. In certain aspects, however, our 
model is simpler than in \citep{Petr}.
In particular, we do not account for 
zooplankton and, therefore, 
do not take into account
possible predator-prey 
interactions in an explicit form.

Let us  assume that the resource supplies $S_k$
can depend on the environmental parameters, for example, temperature $T$:
$S_k=S_k(T)$.   In turn,   $T$ may depend on species abundances, for example, via albedo \citep{Alb}.
We assume, for simplicity,  that this effect is linear:
\begin{equation}
  T=\bar T  + \Delta  T, \quad \Delta T= \sum_{k=1}^N \mu_{kj } x_j,
\end{equation}
where $\mu_{ik}$ are coefficients and $\bar T$ is a reference temperature corresponding to the albedo of the ecosystem environment,
such as the upper ocean,
without the ecosystem influence.  
If the temperature variations $\Delta T$ induced by the species are small, we have
\begin{equation} \label{add1}
  S_k=\bar S_k(\bar T)  +  \Delta S_k + O(\Delta T^2), \quad \Delta S_k=\sum_{k=1}^N b_{kj }(\bar T) x_j, \quad  k=1,..., M, 
\end{equation}
where $b_{kj}= \frac{dS_k(\bar T)}{d\bar T}\mu_{kj}$ .
If all $b_{kj} >0$ we are dealing with purely positive feedback 
(then species abundance increases resources), 
and if  all $b_{kj} < 0$ one has purely negative feedback.

There is, however, an interesting case where some
of the coefficients $b_{kj}$ are positive numbers and others
are negative (mixed feedback).
For mixed feedback a cumulative effect of the climate-ecosystem feedback on the resource supplies may be small since the different terms in $\Delta S_k$ may cancel each other. On other hand, when
the signs of the $b_{jk}$ alternate, but these coefficients are sufficiently large, there may be complicated
large time behavior. We discuss this problem in more detail in Section 5.

There are also possible alternative physical mechanisms 
leading to  relations like (\ref{add1}).  
An important resource for phytoplankton is oxygen \citep{Petr}.   
The production of oxygen is proportional to the phytoplankton 
concentration and depends on temperature $T$.   

Finally,  the extended model takes the form
\begin{equation}
     \frac{dx_i}{dt}=x_i (- r_i  + \phi_i(v) - 
     \sum_{j=1}^N \gamma_{ij} \; x_j), \;\;\;\;\;\; 1 \leq i \leq N,
    \label{HX2E}
     \end{equation}
\begin{equation}
     \frac{dv_j}{dt}=D_j( S_j(x) - v_j)   -  
     \sum_{k=1}^N c_{jk} \; x_k  \; \phi_k(v),  \;\;\;\;\;\; 1 \leq j \leq M,
    \label{HV2E}
     \end{equation}
where
\begin{equation} \label{add1E}
  S_k(x)=\bar S_k  + \sum_{k=1}^N b_{kj }x_j, \quad  k=1,..., M. 
\end{equation}
This model is an approximation of the model with temperature dependent $S$ only up to the terms
of order $\Delta T^2$. 

In the next section we show that under some assumptions this model is well posed.

 \section{General properties of the model}  \label{Genprop}

Let us first describe some sufficient conditions which guarantee 
that systems (\ref{HX11}), (\ref{HV11}), 
(\ref{HX2}), (\ref{HV2}) and (\ref{HX2E}), (\ref{HV2E}) are dissipative 
and have an attractor,
and recall some basic notions. Since there are variations in 
the definition of attractor, for correctness, we follow \citep{HofSch}. 

Let us consider the Cauchy problem defined by 
eqs. (\ref{HX2E}), (\ref{HV2E}) and positive initial 
data in (\ref{Cauch}) below.  
The solution $z(t, z_0)=(x(t), v(t))^{tr}$ with initial data
$z_0=(x(0), v(0)^{tr}$ (where the $^{tr}$ superscript denotes transpose) 
is unique and is defined  
for all $t \ge 0$ (see Lemma 1).  
We then obtain the map $S^t : z_0 \to z(t, z_0)$ 
defining a global semiflow $S^t$, $t \ge 0$ 
in a cone ${\bf C}={\bf R}_{+}^{N+M}$, which 
serves as a phase space.

Given an interval $I \subset R_{+}$ and a set $K \subset {\bf C}$,
 let $K(I) = \{u \in {\bf C}:
u=z(t, z_0), \  t \in I, \  z_0 \in  K \}$.  We denote  $K(t)=K([t,t])$. 
A set $K$ is invariant if $K(t) = K$ for
all $t $, and forward invariant if $K(t) \subset  K$ for all $t > 0$. The omega limit set $\omega(K)$
is the intersection of all $K([t, +\infty))$ over all $t \ge 0$.
Given a forward invariant
set $K$ a subset $B$ of $K$ is an attractor for the semiflow $S^t$ restricted to $K$ provided there exists an open
neighborhood $U\subset K$ of $B$ such that $\omega(U) = B$.

 The stable set $W^s(K)$ of a compact invariant set $K$ is defined
by
$$
W^s(K)=\{z \in {\bf C}:   \omega(z) \ne \emptyset  \ and \  \omega(z) \subset K \}.
$$
In other words, the stable set of $K$ consists of points 
where trajectories enter inside the set $K$, 
and stay in $K$ for large times $t$

The semiflow is dissipative if there exists an attractor $B$ such that $W^s(B) ={\bf C}$. In other
words, for dissipative semiflows the attractor is a minimal invariant set, which attracts all points. If the attractor consists of a single isolated point,
then this point is stable in the standard Liapunov sense.

Define the matrix $\Gamma$ with the entries $\gamma_{ij}$ 
to satisfy one of the following conditions:

\vspace{1ex}

\noindent
{\bf Assumption 1A.} {\em The matrix $ \Gamma$ with the entries 
$\gamma_{ij}$ has a positive dominant diagonal:
 \begin{equation}
      \gamma_{ii} - \sum_{j=1,..., N, j\ne i} |\gamma_{ij}|=\kappa_i >0
      \;\;\;\;\;\; 1 \leq i \leq N.
\label{gamm}
\end{equation}
 }

\noindent
{\bf Assumption 1B.} {\em The matrix $ \Gamma$ has non-negative entries 
 \begin{equation}
      \gamma_{ij}  \ge 0, \quad \gamma_{ii} >0, 
        \;\;\;\;\;\; 1 \leq i,j \leq N.      
\label{gamm2}
\end{equation}
 }

\noindent
Assumption 1A means that species self-regulation is stronger 
than species interaction, while assumption 1B implies that all 
species in our ecosystem compete. 
Let us show that the solutions to (\ref{HX2E}), (\ref{HV2E}) exist, 
and that they are non-negative and bounded.

\vspace{1ex}

\noindent
{\bf Lemma 1.}
{ \em Assume the functions $\phi_j$ satisfy (\ref{MM2a}). Let us consider for eqs. (\ref{HX2E}), (\ref{HV2E}) the Cauchy problem with positive initial data for $x$ and positive initial resources
 \begin{equation}
     x_i(0) > 0,  \quad  v_j(0) >0,  \;\;\;\;\;\;  \forall i \in \{1,\ldots, N\}, 
                               \;\;\; \forall j \in \{1,\ldots, M\}.
\label{Cauch}
\end{equation}
 Then, if either assumption 1A or  1B holds, 
 solutions of this Cauchy problem exist for all $t \ge 0$,  
 are positive and  bounded for large times $t$, that is,
 \begin{equation}
   0 <  x_i(t) < X(t)= X_0+|X_0- \max_{i} x_i(0)| \exp(- \kappa t),  \quad t >0,
\label{Cauch1}
\end{equation}
where $X_0$ is a positive  constant, $\kappa=\gamma X_0$, and 
 \begin{equation}
  0 < v_j(t) < v_j(0)\exp(-D_j t)  + \max_{s \in [0,t]} V_j(s),
\label{Cauch1V}
\end{equation}
where 
$$
V_j(t)=\bar S_j + \bar b_j X(t),    \quad   \bar b_j=\sum_{i=1}^N  (b_{ji})_{+}. 
$$
}

\noindent
{\bf Proof}. 
For a proof, see the Appendix.

\vspace{1ex}

Due to boundness of solutions for large $t$ we then obtain the following corollary.

\vspace{1ex}

\noindent
{\bf Theorem. }
{\em Under the conditions of the previous lemma, system (\ref{HX2E}), (\ref{HV2E}) defines
a global semiflow in the cone ${\bf R}^{N+M}_{+}$.  This semiflow is dissipative and
has a compact attractor}.

\section{Asymptotic approach} \label{Asym}

Our next step is to find asymptotic solutions 
of the system in (\ref{HX2E}) and (\ref{HV2E}), where the $S_k$ 
are defined by (\ref{add1}).  
 We consider the case of large $D_j \gg 1$.  Note that a 
 reduction to a Lotka-Volterra system described below 
 also holds  for bounded $D$ and large resource 
 supplies $S_k \gg 1$.   To simplify the statement, 
 we assume that $D_j=D$ for all $j$. Let us make the change of variables
\begin{equation} \label{assol1}
  v_k= S_k(x) -  \tilde v_k,  \quad \tau=D t.
\end{equation}
System (\ref{HX2E}) and (\ref{HV2E}) then takes the form
\begin{equation}
     \frac{dx_i}{d\tau}=\epsilon x_i (- r_i  + \phi_i(S(x)-\tilde v) -  
     \sum_{j=1}^N \gamma_{ij} \; x_j),
    \label{HX2M}
     \end{equation}
\begin{equation}
     \frac{d\tilde v_j}{d\tau}=-\tilde v_j   -  
     \epsilon U_j(x, \tilde v),
    \label{HV2M}
     \end{equation}
where $\tilde v=(\tilde v_1, ..., \tilde v_M)$,  $\epsilon=D^{-1} \ll 1$ and 
\begin{equation}
      U_j(x, v)=\sum_{k=1}^N c_{jk} \phi_k(S(x) - \tilde v) + 
      \sum_{k=1}^N b_{jk} (\phi_k(S(x) - \tilde v)- 
      r_k -\sum_{kl} \gamma_{kl} x_l). 
    \label{Uv}
     \end{equation}
For small $\epsilon$ equations (\ref{HX2M}) and (\ref{HV2M})  
form a typical system with slow variables $x_j$ and fast
variables $\tilde v$. We can find an asymptotic solution
of (\ref{HV2M}), which has the form
\begin{equation}
     \tilde v_j   =  \epsilon U_j(x, 0)  + O(\epsilon^2).
    \label{HV2As}
     \end{equation}
Finally, for the species abundances $x_i$ we obtain
\begin{equation} \label{LVE}
  \frac{dx_i}{dt}=x_i (\phi_i(S(x)) - r_i -
  \sum_{j=1}^N \gamma_{ij} x_j) +O(\epsilon).  
\end{equation}

\section{Qualitative analysis of large time behavior} \label{Qual}

If the coefficients $b_{lj}$ are small, i.e., the feedback between the resource supply and the climate is weak, then the system (\ref{LVE}) can be simplified by
the Taylor expansion 
$$
\phi_i(S(x))=\phi_i(\bar S) + \sum_{l=1,...,M} \sum_{j=1,...,N} \frac{\partial \phi_i}{\partial S_l}(\bar S) b_{lj} x_j  +...  \; .
$$
Removing terms quadratic in $x_i$, equation (\ref{LVE}) reduces to the Lotka -Volterra system
\begin{equation} \label{LV}
  \frac{dx_i}{dt}=x_i (R_i -\sum_{j=1}^N A_{ij} x_j).  
\end{equation}
where
\begin{equation} \label{LVD}
R_i=\phi_i(\bar S) - r_i,  \quad  A_{ij}=\gamma_{ij} - \sum_{l=1}^M a_{il} b_{lj},
\end{equation}
and
\begin{equation} \label{LVa}
 a_{il} =\frac{\partial \phi_i}{\partial S_l}(\bar S).
\end{equation}

The Lotka-Volterra systems are very well studied 
(see, for example,   \citep{Hof88, Ta96}) and we 
can use these results to help understand how climate warming can
affect ecosystems.  We assume that 1B holds and 
consider the two limiting cases, the
``weak climate'' ({\bf WC}) regime and 
the ``strong climate'' ({\bf SC}) regime.  The {\bf WC} 
case corresponds to weak climate influence, where the 
ecosystem-climate interaction via the coefficients $b_{ik}$ 
is much weaker than the competition effects associated with 
the coefficients $\gamma_{ij}$.  This means that
all  the $|b_{ik}| \ll \gamma$, where $\gamma =||\Gamma||$ is 
a characteristic magnitude of the entries $\gamma_{ij}$. 

In the {\bf SC} case (regime of strong climate influence; 
coefficients determining climate feedback are stronger than 
the coefficients that define species interaction), 
we assume that $|b_{ik}| \gg \gamma$.

In the  {\bf WC} case, system (\ref{LV}) is close to so-called 
competitive systems, which are well studied 
\citep{Hir85, Smith81, Smith91, Zeem, Zeem2}. 
Under some conditions \citep{Hir85, Hof88, Zeem2}  
these systems exhibit no 
stable periodic or chaotic regimes: almost all trajectories 
converge to equilibria, 
which will be investigated in section \ref{eq} for the case 
of a single resource.

Consider the {\bf SC} case.  We set $\gamma_{ij}=0$ for all $i,j$. Then 
equations  (\ref{LV}) represent a Lotka-Volterra system of a special 
structure. An analysis \citep{Hof88}  shows that, for general $R_i$, 
no more than  $M$ species can coexist -- an expression of
the competitive exclusion principle. Mathematically this means 
that if $N >M$ then for some $i$ either  the corresponding 
$x_i(t) \to 0$ or $x_i(t) \to +\infty$ as $t \to +\infty$, 
i.e., the system is not permanent  \citep{Hof88}. However,  if the condition     
\begin{equation} \label{condofmany}
             R_i=\sum_{k=1}^M a_{ik} \theta_k,  \quad \forall i=1,...,N
\end{equation}
for some $\theta_k$ is fulfilled, then it is possible that 
all $N$ species can coexist.     
In this case system (\ref{LV}) can be studied by an idea 
proposed by Volterra \citep{Volt}.  We introduce 
new variables $q_j$, named the qualities of life in \citep{Volt}, 
where $j=1, \ldots, M$. Then eq. (\ref{LV}) reduces to a system 
involving  only the variables $q_j$ \citep{Koz}:
\begin{equation} \label{qsys}
             \frac{dq_j}{dt}=G_j(q),  
\end{equation}
\begin{equation} \label{qsys2}
               G_j(q)=-\theta_i + 
    \sum_{i=1}^N  b_{ji } C_i  \exp(- \sum_{j=1}^M a_{ij} q_j),
\end{equation}
where the $C_i$ are arbitrary positive constants.  
The species abundances $x_i$ can be expressed via $q_j$  by 
$$
x_i= C_i \exp(- \sum_{j=1}^M a_{ij} q_j), \quad  i=1, \ldots, N.
$$ 
Note that  $C_i=x_i(0)$ and therefore the vector field $G(q)$  depends on initial data
and the species number $N$.
So, system  (\ref{qsys}) completely determines dynamics of $x_i$. 

The main results on system (\ref{qsys}) can be outlined  
as follows (see \citep{Koz} for more details). Let $\Omega$ 
be a compact connected domain in ${\bf R}^M$ with a 
smooth boundary, $F(q)$ be a compact $C^1$ smooth
field on $\Omega$, and $\epsilon >0$  be a real number.  
Then there exist a number $N$ and coefficients 
$a_{ij} >0, C_i >0$ and $b_{il}$ such that the 
corresponding field $G$ approximates
$F$ in the domain $\Omega$     in $C^1$-norm  
with accuracy $\epsilon$.   
This approximation result implies that system (\ref{qsys}) 
with $M$ variables $q_j$ can generate all structurally 
stable dynamics in dimension $M$. In particular, 
due to the Theorem on Persistence of 
hyperbolic sets \citep{Ruelle}, system (\ref{qsys}) 
can exhibit all (up to topological orbital equivalences) 
hyperbolic dynamics, including 
periodic and chaotic, including for example, the Smale horseshoe, 
Anosov flows, etc. 

Under condition (\ref{condofmany}) we find that the  
time behavior of solutions of system (\ref{LV})  
depends sharply on $M$. Assume that $a_{ik} > 0$. Note
that this assumption looks natural since
it means that $\phi_i$ increases as a resource supply $S_i$ increases.  

If $M=1$  
it is possible that all $N$ species survive 
in an equilibrium state, and $N$ may be large.  
Although periodic and chaotic trajectories 
are impossible, we can observe multistability
(coexistence of many equilibria). 
   
For $M=2$ and $b_{ik}$ of different signs, system (\ref{LV}) 
can have time periodic solutions   
and for $M >2$ this system can produce time chaotic 
solutions (we can then obtain all possible hyperbolic 
invariant sets of dimension $\le M$).
If all $b_{ik} < 0$ or all $b_{ik} >0$ we have no 
complex behavior for the trajectories and they are convergent. 
Therefore, the most interesting situation arises in the 
biodiversity case when $b_{ik}$ have different signs.    
Finally, we conclude that in the {\bf SC} regime there 
are possible chaotic phenomena and periodic
oscillations if there exist at least 
three resources $v_j$.

In the next subsection we will study  the 
case $M=2$ and we will see that in this case Andronov-Hopf 
bifurcations are possible.

\subsection{Bifurcations, complexity  and biodiversity} \label{bifurc}

If there exists a positive climate-ecosystem feedback, and $b_{ik} >0$,  
then time periodic (for $M >1$) or even chaotic (for $M >2$) behavior,
as well as complicated bifurcations, can occur.   

We consider two cases: $M=1$ (a single resource) and $M=2$,   
and investigate the existence of different bifurcations, 
in  particular, the Andronov-Hopf bifurcations.  If $M=1$  
there are possible saddle-node, pitchfork, and transcritical 
bifurcations, but the Andronov-Hopf does not occur.  
The main climate effect in the case $M=1$ is a destruction 
of the ecosystem under climate forcing that can be described as follows. 
Let us consider a population consisting of $N$  species with 
random parameters, and denote $q=q_1, G=G_1$. We can assume, 
for example,  that the
parameters $a_i$ and $K_i$  in   (\ref{MM5}) and $b_{1i}$  
in (\ref{add1}) are normally distributed random variables. 
The equilibria 
are defined by roots of equation $\theta=G(q)$.  


Let us consider system (\ref{qsys}) for $M=2$.  
Let $(Q_1,Q_2)$ be a steady state for  this system, and
we define a $2 \times 2$ matrix ${\bf M}$ with entries
$$
M_{lj}=\frac{\partial G_l}{\partial q_j}(Q_1, Q_2).
$$
We introduce vectors  $b^{(l)}=col(b_{l1},b_{l2},...., b_{lN})$ and
$$
 E_a(Q)^{(j)}=col(a_{1j} \exp(- a_{11}Q_1 -a_{12} Q_2 ), ..., a_{Nj} 
 \exp(- a_{N1}Q_1 -a_{N2} Q_2 )).
$$ 
  Then we obtain 
$$
  M_{kl}=  {\langle b^{(k)},  E_a(Q)^{(l)}  \rangle }_C, 
  \quad l, k \in \{1,2\}.
$$
An Andronov-Hopf bifurcation occurs if the trace $Tr_M$ 
of the matrix ${\bf M}$
changes its sign as the bifurcation parameter  $b$  goes 
through a critical value $b_c$ and if the determinant 
$det_M$ of ${\bf M}$ is positive  at this critical value. 
Using the notation in (\ref{notat}) we obtain 
\begin{equation} \label{det}
Det_M=   M_{11} M_{22} -M_{12} M_{21}, 
\end{equation}
\begin{equation} \label{Trace}
Tr_M=   {\langle b^{(1)},  E_a(Q)^{(1)}  \rangle }_C   
+  {\langle b^{(2)},  E_a(Q)^{(2)}  \rangle }_C. 
\end{equation}

These relations allow us to see connections between 
bifurcations, feedback, and diversity. First 
let us observe that components of the vectors
$E_a(Q)^j$ are always positive.  Note that if the 
climate influence is absent, then all the components 
of $b^l$ are negative, and it is clear that
$Tr_M$ does not change its sign. Thus in this case 
the Andronov-Hopf bifurcations are absent. 
The same fact holds if all the climate-ecosystem feedbacks
are negative.  For purely  positive or mixed feedbacks 
these bifurcations  are possible under  additional 
conditions. In order to find a biological meaning
of these conditions, we define $\phi_{lj}(C)$ as 
the angles between the vectors $b^{(l)}$ and  $ E_a(Q)^{(j)}$. We then have 
$$
\phi_{lj}(C)=  {\langle b^{(l)},  E_a(Q)^{(j)}  \rangle }_C 
|| b^{(l)}||_C^{-1} || E_a(Q)^{(j)}  ||_C^{-1}.
$$
Then the condition $Det_M >0$ reduces to  
\begin{equation} \label{angledet}
\phi_{11}(C) \phi_{22}(C)  >  \phi_{12}(C) \phi_{21}(C).
\end{equation}
The condition $Tr_M=0$ implies that $\phi_{11}(C)$ 
and $\phi_{22}(C)$ have opposite signs.  
Then (\ref{angledet}) means that
$\phi_{12}(C)$  and $\phi_{21}(C)$ also have   
opposite signs.  If all the species affect the climate 
in a similar manner (the coefficients $b_{kj} $ have the 
same signs) then  all the quantities $\phi_{lj}$ have the 
same sign. Therefore, Andronov-Hopf bifurcations are impossible
in this case.  

We conclude that 
not only feedback positivity but also biodiversity 
and a complex ecosystem structure support complicated 
time periodic behavior. Moreover, all bifurcation conditions
depend on the initial data $C$. From a biological point 
of view, this means that bifurcation effects have a 
"memory", i.e., they depend on the choice of initial data.   

\section{Equilibria} \label{eq}

The aim of this section is to show that the cases 
of negative ({\bf NF}) and positive ({\bf PF})
feedback between climate and
ecosystem are markedly different.  In the {\bf NF} case, 
positive equilibria with many species can exist. 
In the {\bf PF} case, such equilibria vanish 
for some critical feedback level; 
this can be interpreted as a \textit{mass extinction}. 
We compute this critical level.

 On the attractor structure, one can say more for the particular 
case of system (\ref{HX2E}) and (\ref{HV2E}), where we 
have a single resource,  $M=1$. We use  
eqs. (\ref{HX11}), (\ref{HV11}), where  $K_i=K$ and thus
$\phi_i=a_i \phi(v)$,  where $\phi(v)=v/(K+v)$.  
Let us set $\rho_i=r_i/a_i$. 
These quantities
are important characteristics of species.  
The species with smaller $\rho_i$ have a greater chance
to survive. Moreover, using an analogue of \eqref{add1E} 
for the case a single resource $M=1$, we assume 
that $S$ depends on $x$ as follows:
$$
S(x)=\bar S + \sum_{k=1}^N  b_k x_k,
$$
where $b_k$ are the coefficients.

\subsection{Equation for equilibrium resource value}

Moreover, for simplicity,
let us set
\begin{equation}
  \gamma_{ij}= \gamma_i \delta_{ij}, \quad \gamma_i >0.
\label{gam1}
\end{equation}
In this case numerical simulations
show that all trajectories tend to equilibria.  
As was pointed out by V. Kozlov, using the 
theory of decreasing operators and an assumption that
$\phi_i(v)$ increases in the resource $v$ and $D \gg 1$, 
 one can prove this fact by analytic methods 
(a detailed analysis of this question will appear
in future work, since the proof is quite involved). 
The resting points $(\bar x, \bar v)$ of systems (\ref{HX11}) 
and (\ref{HV11}) can be found as follows.

\begin{figure}[t]
\includegraphics[width=1.0\linewidth]{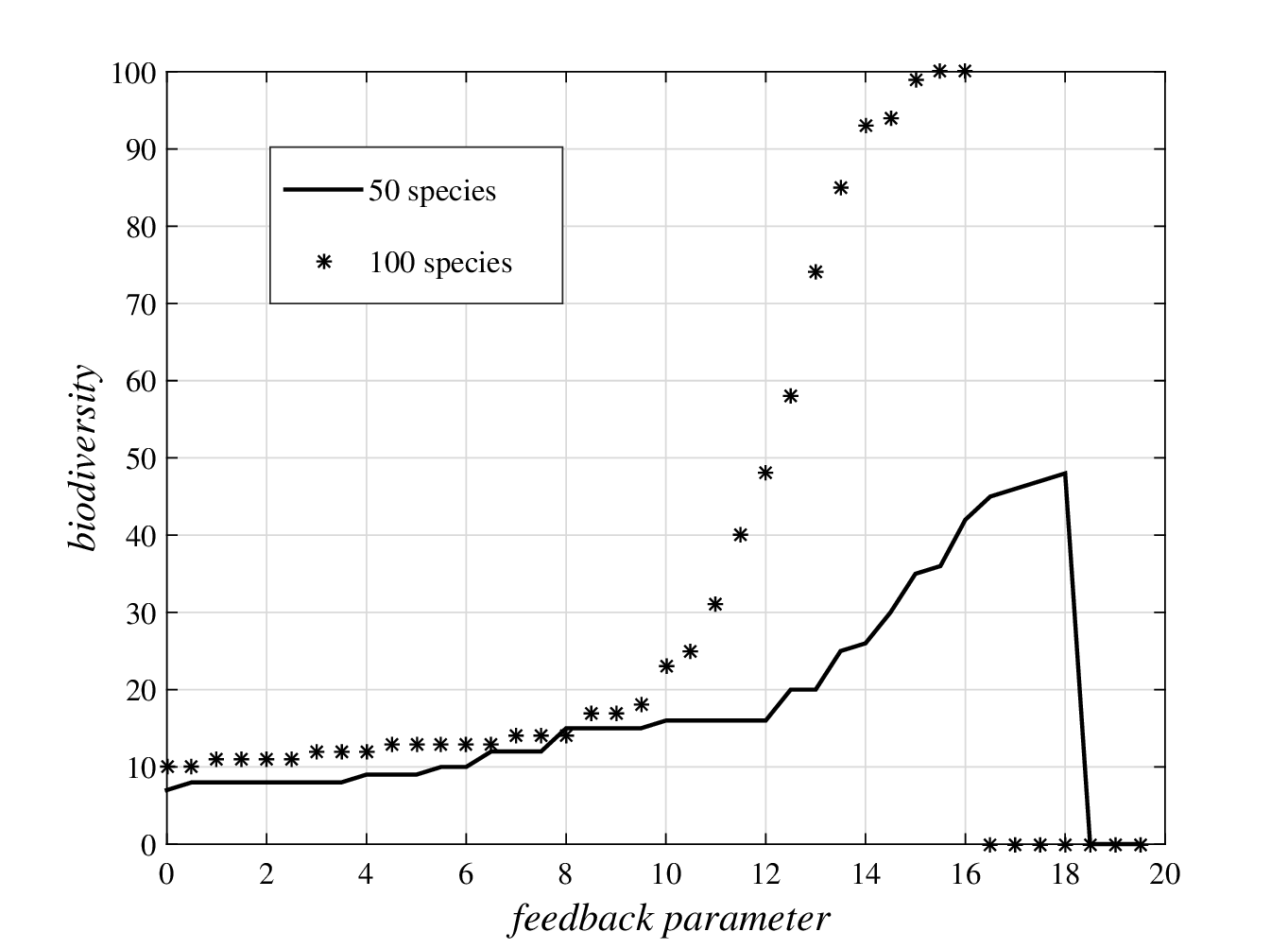} \label{Fig1}
\caption{The graph shows the dependence  
of the biodiversity $N_B$ on the magnitude of positive feedback 
$b_F$, in the case of $N=50$ and $N=100$. 
We see that the biodiversity increases as $b_F$ grows,
but beyond a critical value of the climate-ecosystem feedback
$b_F$, all the species become extinct.  
}
\end{figure}


Setting $dx_i/dt=0$ in (\ref{HX11}), 
we obtain $\bar x_i=a_i \gamma_i^{-1}( \phi(\bar v)-\rho_i)_{+}$. This gives
the following nonlinear equation for $\bar v$:  
\begin{equation} \label{eqbr}
 D(\bar S -\bar v)=G(v),
\end{equation}
where
\begin{equation} \label{GV}
G(v)= \sum_{j=1}^N a_j \gamma_j^{-1}(c_j  a_j   \phi(v) -    
b_j )(\phi(\bar v)- \rho_j)_{+}.
\end{equation}

We have obtained a complicated equation with non-smooth 
nonlinearities. An important characteristic
of the solutions $\bar v$ is $N_B(\bar v)$, the number 
of positive $\bar x_j(\bar v)$ involved
in the sum in the right hand side of  (\ref{GV}).  
The number $N_B$ can be interpreted as biodiversity.

Note  that, for any $N$, 
in the {\bf NF} case a solution 
$\bar v$ with $\bar v \in (0, \bar S)$ always exists
under the following condition:
\begin{equation} \label{CE}
\phi(S) > \rho_0=\min_j \rho_j.
\end{equation}
Indeed, observe that
$D(\bar S -v)$ is a decreasing function of $v$, while $G(v)$ 
is increasing. The solution
$\bar v$ is given by an intersection of the curve  $G(v)$ 
and the right line $D(S-v)$, which exists
if (\ref{CE}) holds.

Moreover, the same geometrical argument shows that the 
resource $\bar v$  is an increasing function
of $b_j$. Therefore, in the case of negative feedback 
the biodiversity $N_B$  is larger (if a solution $\bar v >0$ exists).  
However, for negative $b_j$ that are too large, 
the positive solution $\bar v$ does not exist.   

Consider a large  ecosystem with random parameters $\rho_j$.  We suppose
that $N~\gg~1$ and $\rho_j$ are selected randomly according to a 
distribution with 
probability density function $\xi(\rho)$, which is positive on some open interval 
$I_{\rho}=(R_{0}, \ R_{1})$. 

\vspace{1ex}

\noindent
{\bf Assertion}.  {\em 
Consider the case of  negative feedback ($b_j \ge 0$ ). If
\begin{equation} \label{EEE}
\phi(\bar S) > R_{0},
\end{equation}
 then for any $N$ there exists a positive solution  $\bar v(N)$  of eq. 
(\ref{eqbr}) with biodiversity $N_B(\bar v)$ such that $N_B \to \infty$ as $N \to \infty$.  
}

\vspace{1ex}

\noindent
{\bf Proof}.  The existence of solutions is obvious from geometrical arguments 
(see remarks on the monotonicity of $D(\bar S- v)$ and $G(v)$ above).  To show that
$N_B$ is large for $N~\gg~1$, we observe that for any fixed  $\rho_{0}$ and $\rho_{1}$ 
such that $R_{0} < R_{1}$, the  
interval $(R_{0}, R_{1})$ contains $N_c$ points 
$\rho_j$, with $N_c \to  +\infty$ as $N \to +\infty$.
For large $N$ we seek a solution of (\ref{eqbr})  
in the form $\bar v=S - w$, where $0< w~\ll~1$. 
Since (\ref{EEE}) holds, such a solution exists.  
The number $N_b$ approximately equals
the number $N_c$ for $R_{1}=\phi(\bar c) \approx \phi(\bar S)$, 
and the assertion is proved.

\vspace{0.2cm}

In the {\bf PF} case this assertion, in general, does not hold. Using the arguments
from the proof, we note that all species die  if the following relation holds:
\begin{equation} \label{GV}
R_0 \sum_{j=1}^N a_j^2 \gamma_j^{-1}c_j     <   \sum_{j=1}^N 
 b_j a_j \gamma_j^{-1} .
\end{equation}
This relation shows that mass extinction inevitably 
arises if the $b_j >0$ are sufficiently large.

Results on a numerical solution of equation (\ref{eqbr}) are discussed below. 
They confirm that
mass extinctions are possible as the feedback magnitude increases.

\subsection{Numerical results} \label{Numres}


In the general case  equation \eqref{eqbr} for equilibria 
can be resolved numerically for $N=50$.  We choose the 
coefficients in equation (\ref{eqbr})  as follows. The positive 
coefficients $a_i$ are random numbers subject to log-normal 
distributions.  This means that $\ln(a_i)$ are distributed normally, 
$\ln a_i \in {\bf N}(E_a, s_a)$, where $E_a$ is the mean and 
$s_a$ is the deviation.   The same distribution is taken for 
$c_i$, with the parameters $E_c$ and $s_c$.

We assume that the $R_i$ and $b_i$ are distributed normally, namely,
$R_i \in {\bf N}(R_0, s_R)$ and $b_i=b_F \beta_i$, where
 $\beta_i  \in {\bf N}(b_0, s_b)$, and $b_F$ is the magnitude 
 of the feedback level. 
The other parameters were taken as follows: 
$D=1, K=2, \bar S=10, E_a=1,  s_a=0.1, E_b=1, s_b=0.3,  R_0=
0.7,  S_R=0.05$ and $\gamma_i=1$.   

The results are shown by Fig.1\label {Fig1}. 
Comparison of the two plots shows that when the number 
of species increases, so does the likelihood  of 
a sharp drop in species number
as the climate changes and feedback processes grow stronger. 
These findings are consistent with analytical 
results. Biodiversity grows
with the feedback parameter $b_F$, until at some critical level  
we observe a  mass extinction.

\section{Conclusions}

In this paper, a resource  model for a system of many 
coexisting species is proposed. It
is a generalization of the well known model in \citep{HuWe99}, 
takes into account  
species self-regulation and a dependence on the environment,  
and is the first model of an ecosystem with many species
and feedback which couples 
climate and population dynamics. 
Such conceptual models describe a simple and easily understandable 
mechanism for resource competition. 
For the case of fixed parameters, a general assertion on 
attractor existence for this model is proved.  
One of the sufficient conditions for the existence of an attractor 
is that
species self-regulation is stronger than species competition.  

 Climate-ecosystem feedbacks 
are an important problem in terms of uncertainty in predictions 
and modeling future climate change. The proposed model allows us not only 
to investigate climate-ecosystem feedbacks for large ecosystems, 
but also to show that
coexistence of many species feeding on a few resources is possible. 
In the case of positive feedback in the ecosystem-climate interaction,
the numerical results show a possibility of catastrophic bifurcations,  
when all (or almost all) species become extinct under the impact of climate warming. 
The ecosystem biodiversity increases with
the magnitude of positive feedback $b_F$, but at some critical 
level of feedback, a mass extinction occurs.
For negative climate-ecosystem feedback we observe smaller  
biomass and biodiversity values,
but we do not observe catastrophes. Note that in the contemporary 
world, human impact on the climate system can possibly lead to positive feedback
in the above context.

To investigate more complicated situations, where complex 
dynamics may be possible, we have considered the case of just
a few resources. 
We find asymptotic solutions for the case of a large resource turnover. 
This allows us to reduce
this system to the Lotka-Volterra model, which is well studied.  
The existence of two sharply different
regimes of ecosystem behavior is proven: the weak climate 
regime ({\bf WC}), and the strong climate regime ({\bf SC}).  
This behavior depends on 
 a parameter 
that determines the intensity of ecosystem-climate interactions. 
Note that this analytical result 
is consistent with experimental data \citep{Paleo}, where it is shown 
that two distinct regimes of extinction dynamic are present in the major
marine plankton group.   
Results in \citep{Paleo} suggest that the
dominant, primary controls on extinction were abiotic (environmental), 
which corresponds to the {\bf SC} case.

In the {\bf SC} case we do not observe
complicated dynamical effects when 
the ecosystem -- climate 
interactions involve only negative or only positive
feedback loops. 
However, if the ecosystem -- climate interaction involves terms 
of different signs, then there are possible 
Andronov-Hopf bifurcations, time periodic behavior for the case 
of two resources, and chaotic behavior for more than  three resources. 
We conclude that 
not only feedback positivity, but also biodiversity and a 
complex ecosystem structure (when different species affect 
climate differently creating 
positive and negative feedback ecosystem -- climate loops) 
support complicated temporal dynamics of the ecosystem.    

For the case of a single resource the ecosystem equilibria can 
be described implicitly.
We find these equilibria by a nonlinear equation for the 
equilibrium resource level. We show that, due to self-limitation 
effects, the system can support equilibria
with a number of species sharing the same single resource.

\vspace{3ex}

\section*{Acknowledgments}

The authors are grateful to the anonymous reviewers and the guest editor for interesting remarks and comments, which 
significantly improved the paper.  We are thankful to Prof. M. L. Zeeman (Bowdoin College) for 
very helpful comments on the systems we study here. 
We also express our gratitude to Prof. V. Kozlov (Linkoping University)
for his help. 

This study was funded by RFBR, through 
research projects No.16-34-00733 mol\_a and No.16-31-60070 mol\_a\_dk. 
We gratefully acknowledge support from the Government of the Russian Federation through mega-grant 074-U01, as well as from the Division of Mathematical Sciences and the Division of Polar Programs at the U.S. National Science Foundation (NSF) through Grants DMS-0940249 and DMS-1413454. We are also grateful for support from the Office of Naval Research (ONR) through Grant N00014-13-10291.
Finally, we would like to thank the NSF Math Climate Research Network (MCRN) as well for their support of this work. 

\section*{Appendix}

We state here the proof of Lemma 1. The proof  proceeds in the following steps. 

{\em Step 1}. Positivity of the $x_i$ follows from the fact that the $i$-th right hand side of system (\ref{HX2}) is proportional to
$x_i$, thus, $x_i(t)=x_i(0)\exp(\xi_i(t))$, where $\xi_i$ is a function.

{\em Step 2}. Let us prove that $v_j(t) >0$. Assume that this fact is violated.
Then  there
exists an index $j_0$ and a time $t_0 > 0$ such that
 \begin{equation}
   v_{j_0}(t_0)=0,  \quad \frac{dv_{j_0}}{dt} \le 0, 
\quad v_j(t_0) \ge 0, \quad \forall  \ j.
\label{Cauch3}
\end{equation}
Condition  (\ref{MM2b}) entails the term $ \sum_{k=1}^N c_{jk} \; x_k  \; \phi_k(v)$ equaling zero.  
Then we substitute these inequalities into the $j_0$-th equation (\ref{HV2E}) and obtain a contradiction.

{\em Step 3}.  Let us prove estimate (\ref{Cauch1}). First let us suppose that assumption 1B is satisfied.
Let $E(t)=\max \{x_1(t), ..., x_N(t)\}$. Let us estimate $dE/dt$ for large $E$.  Let $i_0(t)$ be an index
such that $E(t)=x_{i_0}(t)$.
According to (\ref{MM2a}) the $\phi_i$ are uniformly bounded by $C_+$.
Therefore  within any open interval $I_{i_0}$, where $i_0$ is fixed, one has
\begin{equation} \label{ineq1}
\frac{dE}{dt} \le E R_{i_0},  \quad  R_{i_0} \le C_{+}  - \gamma  E(t),
\end{equation}
where  $\gamma=\min_{i} \gamma_{ii} >0$ due to  assumption (\ref{gamm}) on $\Gamma$.

In the case 1A we note that
$$
\sum_{j=1}^N \gamma_{i_0 j} x_j \ge \gamma_{ii} E -  \sum_{j \ne i_0} |\gamma_{i_0 j}| x_j
\ge \kappa E, 
$$
and we have an inequality analogous to (\ref{ineq1}):  
\begin{equation} \label{ineq1a}
\frac{dE}{dt} \le E R_{i_0},  \quad  R_{i_0} \le C_{+}  - \kappa  E(t).
\end{equation}
 
Note that the sequence of intervals $I_{i_0}$ is not bounded and these intervals cover all
${\bf R}_{+}$ since,  according to the Lemma, the solutions exist for all $t>0$.

Inequality (\ref{ineq1}) implies
that 
$
E(t) \le X(t),  
$
where $X(t)$ is the solution to the Cauchy problem 
\begin{equation} \label{XC}
\frac{dX}{dt}=X (C_{+} - \gamma_0 X), \quad X(0)=\max_i x_i(0), 
\end{equation}
where $\gamma_0$ equals $\gamma$ in the case 1B and $\kappa$ in the case 1A.  
Let $X_0=C_{+}/\gamma_0$. If $X(0) < X_0$, then  equation (\ref{XC}) shows that
$X(t) \le X_0$ for all $t$ and (\ref{Cauch1}) follows.  
If $X(0) > X_0$, then  equation (\ref{XC}) shows that
$X(t) > X_0$ for all $t$.  By the change of variables 
$\tilde X=X-X_0$ we obtain that $\tilde X >0$ and thus
$$
\frac{d\tilde X}{dt}=-\gamma_0(X_0 +\tilde X) \tilde X \le -\gamma_0 X_0 \tilde X,
$$
which implies $\tilde X(t) \le \tilde X(0) \exp(-\gamma_0 X_0 t)$, and 
 we obtain (\ref{Cauch1}).

{\em Step 4}.  Having (\ref{Cauch1}),  we can prove  (\ref{Cauch1V}). 
Indeed, using the non-negativity of the $c_{jk}$ and $\phi_k$, one obtains
$$
\frac{dv_j}{dt} \le D_j(S_j(x(t)) -v_j).
$$
Therefore, 
$$
v_j(t)=\exp(-D_j t)( v_j(0)  +\int_0^t  S_j(x(s)) \exp(D_j s) ds)
$$
which yields
$$
v_j(t)  \le \exp(-D_j t) v_j(0)+ max_{s \in [0,t]}
 S_j(x(s)).
$$
Here $S_j(x(t)) \le  \bar S_j + \bar b_j X(t)$.
These two last inequalities imply $v_j(t) \le V_j(t)$, which completes the proof.

\section*{\bibliographystyle{elsarticle-harv}}

\end{document}